\documentclass[preprint,aps,nofootinbib]{revtex4}

\usepackage{graphicx}

\usepackage{amsmath}
\usepackage{amsfonts}
\usepackage{amssymb}

\usepackage{epstopdf}

\begin{document}


\title{
Probing the Stop Sector of the MSSM with the Higgs Boson at the LHC}
\vspace*{2cm}

\author{
\vspace{1cm} 
Radovan Derm\'\i \v sek$^{\, a}$ and  Ian Low$^{\, b}$}

\affiliation{
\vspace*{.5cm}
\mbox{$^a$School of Natural Sciences, Institute for Advanced Study, Princeton, NJ 08540} \\ 
\mbox{$^b$Department of Physics and Astronomy, University of California, Irvine, CA 92697}
\vspace*{1cm}
}

\begin{abstract}
\vspace*{0.5cm}

We propose using the lightest CP-even Higgs boson in the minimal supersymmetric 
standard model (MSSM) to probe the stop sector. Unlike measuring stop masses 
in production/decay processes which requires knowledge of masses and mixing angles of other 
superparticles, our strategy depends little on supersymmetric parameters other than
those in the stop sector in a large region of the parameter space. We show that 
measurements of the
Higgs mass and the production rate in the gluon fusion channel, the dominant channel 
at the LHC, allow for determination of two parameters in the stop mass-squared matrix,
including the off-diagonal mixing term. This proposal is very effective when stops are light 
and their mixing is large, which coincides with the region where the electroweak symmetry breaking 
is minimally fine tuned.
We also argue that a lightest CP-even Higgs mass in the upper range of allowed values
and a production rate significantly smaller than the rate predicted in the standard model would be 
difficult to reconcile within the MSSM, except in extreme corners of the parameter 
space.

\end{abstract}


\maketitle

\section{Introduction}
\label{sec:introduction}

Supersymmetry (SUSY) is usually considered the leading candidate for physics beyond the standard model.
Among many virtues of SUSY, perhaps the most prominent ones are the stabilization of the electroweak
scale up to very high energies such as the grand unification scale and the possibility of radiatively
driven  electroweak symmetry breaking (EWSB). 
However, neither the Higgs boson nor
any superpartners have been found in collider experiments so far, and it is  
discomforting to realize that majority of natural parameter space of MSSM has been ruled out by current 
experimental limits on the Higgs mass~\cite{LHWG}, leaving us with the parameter space where EWSB is achieved with fine tuning of soft SUSY breaking parameters at a few percent level.~\footnote{For recent discussion of fine tuning in EWSB see e.g. Refs.~\cite{Dermisek:2005ar,Choi:2005hd,Kitano:2005wc,Dermisek:2006ey,Dermisek:2006qj}.}

The EWSB and the mass of the Higgs boson in the MSSM are 
tightly connected with the stop sector: stop mass squared parameters, $m_{\tilde{t}_L}^2$ and $m_{\tilde{t}_R}^2$, and the mixing, $X_t = A_t -\mu/\tan\beta$, where $A_t$ is the top soft trilinear coupling, $\mu$ is the supersymmetric Higgs mass and $\tan \beta = v_u/v_d$ is the ratio of the vacuum expectation
values of up-type and down-type Higgs bosons. These parameters enter the calculation of physical stop masses, $m_{\tilde{t}_1}$ and $m_{\tilde{t}_2}$, which is what we measure in experiments. 
Information about the mixing is not given from mass eigenstates.
The splitting between 
$m_{\tilde{t}_1}$ and $m_{\tilde{t}_2}$ can originate either from the difference between $m_{\tilde{t}_L}^2$ and $m_{\tilde{t}_R}^2$ or from large mixing. However, the mixing in the stop sector is crucial for the Higgs boson mass.
In the MSSM the mass of the lightest CP-even Higgs boson is bounded at the tree-level by the $Z$ boson mass,
$m_h \le m_Z |\cos 2\beta |$. 
In order to lift the Higgs mass above the LEP limit $m_h\ge 114$
GeV, radiative corrections from stops are required to be large, which then implies either  stop masses 
heavier than about 900 GeV for moderate mixing, or large stop mixing for fairly light stop masses.
Indeed the region of large mixing, $X_t/m_{\tilde{t}_{L,R}}\simeq \pm 2$, and stop masses $m_{\tilde{t}_L} \simeq m_{\tilde{t}_R} \simeq 300$ GeV minimizes the fine tuning of EWSB while satisfying the limit on the Higgs mass.
There has been some effort to realize the large mixing scenario in models, see e.g.
Refs.~\cite{Choi:2005hd,Kitano:2005wc,Dermisek:2006ey,Dermisek:2006qj}, 
in order to address the naturalness issue of the MSSM.
It goes without saying that
determining parameters of the stop sector
precisely in collider experiments will be of great importance for understanding the EWSB, the Higgs mass and the internal consistency of the MSSM.~\footnote{It is important to note that the discussion in this paragraph is specific to the MSSM. In models with more complicated Higgs sector, the mass of the Higgs boson can receive additional contributions or the 114 GeV limit on the Higgs mass might not apply due to modified Higgs decays.
See e.g.~Ref.~\cite{Dermisek:2005ar} for related discussion.}

So how does one measure stop masses and the mixing angle? This is a simple question without a simple answer.
In the MSSM with R-parity the lightest supersymmetric particle (LSP) is stable. In a 
large class of models the lightest neutralino 
is (or can be) the LSP and a good candidate for dark matter. 
In collider experiments the lightest neutralino (being stable and  
electrically neutral)
will escape direct detection and result in events with missing transverse energy ($E_T$). 
Due to R-parity superparticles need to be pair-produced and they 
eventually cascade-decay into the LSP plus standard model particles. Thus a typical event for
the production and decay of superparticles is multi-jet and multi-lepton with large missing $E_T$. 
In the end the stop,
if produced, is never directly observed in collider detectors.
Any reconstruction of masses and mixing angle in the production/decay processes 
has to rely on the visible decay product and  missing $E_T$. 

At the Large Hadron Collider (LHC) various reasons complicate the measurement of masses and mixing angle in 
the production/decay process. First, the large missing $E_T$ makes event-by-event reconstruction of
masses impossible; one has to resort to measuring kinematic endpoints and edges of invariant mass distributions
of final particles. The position of such endpoints and edges is sensitive to masses of all particles 
involved in the decay chain, including the LSP which escapes detection. Second, at hadron colliders it is the 
partons 
inside the proton that collide with each other and the center-of-mass energy is not a known quantity. 
Thus there is no kinematic constraint to impose in the longitudinal direction of the collider. Third, because
of long decay chains of SUSY particles there are usually many jets and leptons in the final state, leading
to large combinatoric factors. Previous studies \cite{Weiglein:2004hn} showed that in the end it is quite
a complicated and elaborate analysis to extract mass parameters in the production/decay processes, and 
the outcome crucially depends on knowing the mass and nature of other particles in the 
decay chain such as charginos and neutralinos. For stops, there is an added layer of complexity because
decays of stops sometimes involve top quarks, which require extra efforts to identify.

In this paper we propose an approach, complementary to studying the production/decay processes of stops, that does
not require prior knowledge of masses and mixing angles of other superparticles. The proposal is to use 
properties of 
the lightest CP-even Higgs boson in the MSSM to extract parameters in the stop sector. 
At the LHC the Higgs boson is produced dominantly through the gluon fusion process $gg \to h$ and subsequently
decays into other standard model particles. By measuring the invariant mass of the decay products it is possible
to determine the Higgs mass precisely. As it turns out in the MSSM both the gluon fusion production rate
and the Higgs mass are sensitive only to parameters in the stop sector and not to  masses and mixing of other superpartners. The only
exception to this is the large $\tan \beta$ region where contributions from sbottom  sector to both the Higgs mass and  
the gluon fusion rate can be significant. 
Furthermore, we will demonstrate that, if the variables are chosen appropriately, 
the Higgs mass and the gluon fusion rate depend on only two out of the three parameters in the mass
matrix; the dependence on the third parameter is negligible in a significant region of the parameter space. 
Therefore with two measurements in the Higgs 
sector we are able to extract two parameters in the stop mass-squared matrix, including the
mixing term $X_t$.

There is also an interesting possibility that the two measurements (the Higgs mass and the gluon production rate) 
would point to mutually inconsistent
values of stop masses and mixing, even 
after taking into account the current (large) estimates of experimental and theoretical errors.
This is
 the case for a large Higgs mass $m_h \agt 130$ GeV and a 
significantly reduced production rate in the gluon fusion channel. Even though, taken separately, 
these two measurements are perfectly allowed in the MSSM, we will
argue that the combined scenario is very difficult to reconcile except in some 
extreme (insane) corners of the parameter space.
Finally, in every SUSY breaking scenario in which $m_{\tilde{t}_L}^2$ and $m_{\tilde{t}_R}^2$
are related to each other in any specific way, and, in addition, parameters in the sbottom sector are related
to parameters in the stop sector, our procedure  can be used to fix the parameters of the model, 
or it could possibly rule out the scenario if the measured value of the Higgs mass and the gluon production rate
are impossible to satisfy for any choice of parameters.

\section{The stop sector in MSSM}
\label{stopinmssm}

The stop mass-squared matrix in the MSSM in the flavor basis $(\tilde{t}_L, \tilde{t}_R)$ is 
given by \cite{Djouadi:2005gj}
\begin{equation}
\label{stopmass}
M^2_{\tilde{t}} = \left( \begin{array}{cc} 
   m^2_{\tilde{t}_L} + m_t^2 + D_L^t & m_t X_t \\
    m_t X_t &  m^2_{\tilde{t}_R} + m_t^2 + D_R^t
                         \end{array} \right),
\end{equation}
where
\begin{eqnarray}
D_L^t &=& \left(\frac12-\frac23 s_w^2\right) m_Z^2 \cos 2\beta , \\
D_R^t &=&  \frac23 s_w^2 m_Z^2 \cos 2\beta , \\
X_t &=& A_t - \frac{\mu}{\tan\beta} .
\end{eqnarray}
In the above $s_w$ is the sine of Weinberg angle. From Eq.~(\ref{stopmass}) we see that
there are four free parameters in the stop mass matrix: $\tan\beta$ (through the dependence on
$\cos 2\beta$), $m^2_{\tilde{t}_L}$, $m^2_{\tilde{t}_R}$, and $X_t$. Nevertheless, the 
dependence on $\tan\beta$ is rather weak because $m_Z^2 \ll m_t^2$. Furthermore,
the mass of the lightest CP-even Higgs boson
in MSSM is insensitive to $\tan\beta$ once $\tan\beta \agt 10$, in which case $\cos2\beta \sim 1$.
In this way neither the Higgs mass nor the stop mass-squared matrix is dependent on $\tan\beta$.
On the other hand, the off-diagonal mixing in the sbottom mass matrix,
\begin{equation}
m_b X_b = m_b(A_b - \mu\tan\beta),
\end{equation}
becomes substantial when $\tan\beta \sim m_t/m_b$ and the supersymmetric Higgs mass $\mu$ is
large simultaneously. In this situation the sbottom contribution to both the Higgs mass and the production
rate in the gluon fusion channel could be significant \cite{Brignole:2002bz,Djouadi:1998az}. 
 Therefore the region of parameter space we would like to focus on in this paper
is:
\begin{itemize}

\item $10 \; \lesssim \; \tan \beta \; \lesssim m_t/m_b$,

\item $|m_b \, \mu \tan \beta| \; \lesssim \; m^2_{\tilde{b}_L}, \ m^2_{\tilde{b}_R}$,

\end{itemize}
for which our strategy will not depend on SUSY parameters other than those in the stop sector.
In this case the stop mass matrix is controlled by three parameters $m^2_{\tilde{t}_L}$,
$m^2_{\tilde{t}_R}$, and $X_t$.
In addition we will be interested in the so-called ``decoupling limit''\cite{Haber:1995be},
in which the lightest CP-even Higgs, $h$,
is standard model-like in that its couplings to quarks and leptons approach the standard model values, 
and all other Higgs bosons in the MSSM are much heavier than $h$ and roughly degenerate. 

\section{Stops and the gluon fusion production}
\label{glue}

\begin{figure}[b]
\includegraphics[scale=0.85,angle=0]{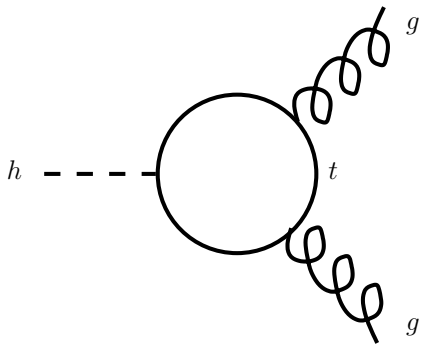}
\caption{\label{fig1}{Gluon fusion production of the Higgs boson in the standard model.}}
\end{figure}

At hadron colliders the dominant production mechanism for the Higgs boson is the gluon fusion production
\cite{Georgi:1977gs,Ellis:1979jy,Djouadi:2005gi}. The contributing Feynman diagram is shown in Fig.~\ref{fig1}, 
in which it is the
top quark running in the loop. The gluon fusion production, being a loop induced process, is very sensitive
to new physics, especially to any new colored particle which couples to the Higgs significantly. In the MSSM
there is only one such particle, the stop, whereas all the other colored superparticles have much
smaller coupling to the lightest CP-even Higgs due to small Yukawa couplings. 
Therefore in the MSSM the gluon fusion production rate of the lightest CP-even Higgs boson probes the stop
sector and is insensitive to other parts of the spectrum.\footnote{The exception is, as commented earlier, 
the contribution from the sbottom sector
for very large
$\tan\beta \sim m_t/m_b$ 
and small sbottom masses~\cite{Djouadi:1998az}.}

Obviously the gluon fusion production rate is directly proportional to the decay rate of $h\to gg$, for
which the stop contribution at one-loop level has been computed \cite{Djouadi:2005gj}. 
The analytic expression, including
the top quark contribution, is
\begin{equation}
\label{hggcorr}
\Gamma(h \to gg) = \frac{G_F \alpha_s^2 m_h^3}{36 \sqrt{2}\pi^3} 
       \left| \frac34 A_{\frac12}^h(\tau_t) + 
      \sum_{i=1,2}\frac34 \frac{g_{h\tilde{t}_i\tilde{t}_i}}{m_{\tilde{t}_i}^2}
               A_{0}^h(\tau_{\tilde{t}_i})\right|^2 ,
\end{equation}
where $\tau_i = m_h^2/(4m_i^2)$ and the form factors are
\begin{eqnarray}
A_{\frac12}^h(\tau) &=& \frac2{\tau^2} [ \tau + (\tau - 1)f(\tau)]\ , \\
A_0^h(\tau) &=& -\frac{1}{\tau^2}\left[ \tau - f(\tau) \right] \ , \\
\label{ftau}
f(\tau) &=& \left\{ \begin{array}{lc}
       \displaystyle     \arcsin^2 \sqrt{\tau} & \quad \tau \le 1 \\
  \displaystyle -\frac14 \left[ \log 
    \frac{1+\sqrt{1-\tau^{-1}}}{1-\sqrt{1-\tau^{-1}}} - i \pi \right]^2 
      & \quad \tau > 1 
         \end{array} \right. .
\end{eqnarray}
Furthermore, $g_{h\tilde{t}_i\tilde{t}_i}$ is the coupling of 
the lightest CP-even Higgs boson to stop mass eigenstates, normalized to $2(\sqrt{2}G_F)^{1/2}$,
\begin{eqnarray}
g_{h\tilde{t}_1\tilde{t}_1} &=&  m_Z^2 \cos 2\beta
   \left(\frac12 \cos^2 \theta_t-\frac23 s_w^2 \cos 2\theta_t\right)  
           +  m_t^2 - \frac12 m_t X_t  \sin 2\theta_t , \\
g_{h\tilde{t}_2\tilde{t}_2} &=&  m_Z^2 \cos 2\beta
   \left(\frac12 \sin^2 \theta_t+\frac23 s_w^2 \cos 2\theta_t\right) 
           +  m_t^2 + \frac12 m_t X_t  \sin 2\theta_t ,
\end{eqnarray}
where $\theta_t$ is the mixing angle between the flavor basis and mass eigenbasis,
\begin{equation}
\label{thetat}
\sin 2\theta_t = -\frac{2m_t X_t}{m_{\tilde{t}_1}^2-m_{\tilde{t}_2}^2} , \quad
\cos 2\theta_t = \frac{m_{\tilde{t}_L}^2+D_L^t-m_{\tilde{t}_R}^2-D_L^t}{m_{\tilde{t}_1}^2-m_{\tilde{t}_2}^2} , 
\end{equation}
such that 
\begin{equation}
\left( \begin{array}{cc}
   \cos \theta_t & -\sin \theta_t \\
   \sin \theta_t & \cos \theta_t 
       \end{array}   \right) 
      M_{\tilde{t}}^2 
\left( \begin{array}{cc}
   \cos \theta_t & \sin \theta_t \\
   -\sin \theta_t & \cos \theta_t 
       \end{array}   \right)  = 
\left( \begin{array}{cc}
     m_{\tilde{t}_1}^2 & 0 \\
     0 &   m_{\tilde{t}_2}^2
       \end{array}   \right).
\end{equation}
The form factors $A_0^h(\tau)$ and $A_{\frac12}^h(\tau)$ approach 4/3 and 1/3, respectively, for 
$\tau_i=m_h^2/(4m_i^2)\to 0$. For $m_h \sim 120$ GeV, $m_t = 172$ GeV, and $m_{\tilde{t}} \sim 200$
GeV, one can check that $\tau\to 0$ is a good approximation for the form factors. 

The $\tau_i
\to 0$ limit is equivalent to approximating the one-loop diagram in Fig.~\ref{fig1} by a 
dimension-five local operator $(h/v) G_{\mu\nu}^a G^{a\,\mu\nu}$, whose coefficient has long been
known to be related to the QCD one-loop beta function \cite{Ellis:1975ap,Shifman:1979eb}. 
If we turn on a background
Higgs field $h$ and consider
the squark threshold effect for the running of one-loop beta function of QCD, neglecting other contributions
for now, we get
\begin{eqnarray}
- \frac1{4g^2(\mu_{\rm r})} G_{\mu\nu}^a G^{a\,\mu\nu} &=& 
  -\frac14 \left( \frac1{g^2(\Lambda)} -   
 \frac{b_3^{{\rm UV}}}{16\pi^2} \log \frac{\Lambda^2}{\mu_{\rm r}^2}-
 \sum_{i=1,2} \frac{ b_3^{(0)}}{16\pi^2} \log \frac{m_{\tilde{t}_i}^2(h)}{\mu_{\rm r}^2} 
   -\cdots \right) G_{\mu\nu}^a G^{a\,\mu\nu} \nonumber \\
&=& -\frac14\left( -  \frac{ b_3^{(0)}}{16\pi^2} \log \frac{\det M_{\tilde{t}}^2(h)}{\mu_{\rm r}^2} 
      -\cdots\right) G_{\mu\nu}^a G^{a\,\mu\nu},
\end{eqnarray}
where $b_3^{(0)} = 1/6$ \cite{Jones:1981we}. Expanding $\det M_{\tilde{t}}^2(h)$ in the presence
of the background Higgs field $h$ with respect to $\langle h \rangle = v/\sqrt{2}$,
 one immediately obtains the dimension-five operator 
$(h/v)G_{\mu\nu}^a G^{a\,\mu\nu}$, whose coefficient is essentially determined by the quantity
\begin{equation}
\left.v \frac{\partial}{\partial h} \log \det M_{\tilde{t}}^2(h) \right|_{h=v/\sqrt{2}}.
\end{equation}
In fact, it is straightforward to verify in Eq.~(\ref{hggcorr}) that in the limit $\tau_i \to 0$
the stop contribution to the decay width $\Gamma(h\to gg)$ is controlled by 
\begin{eqnarray}
\label{stophgg}
 \sum_{i=1,2} \frac{g_{h\tilde{t}_i\tilde{t}_i}}{m_{\tilde{t}_i}^2}
  &=& \frac{m_{\tilde{t}_1}^2 g_{h\tilde{t}_2\tilde{t}_2}+ m_{\tilde{t}_2}^2 g_{h\tilde{t}_1\tilde{t}_1}}
        {\det M_{\tilde{t}}^2 }  , \nonumber \\
  &=& \frac12 \left.v \frac{\partial}{\partial h} \log \det M_{\tilde{t}}^2(h) \right|_{h=v/\sqrt{2}}.
\end{eqnarray}
\begin{figure}[t]
\includegraphics[scale=1.]{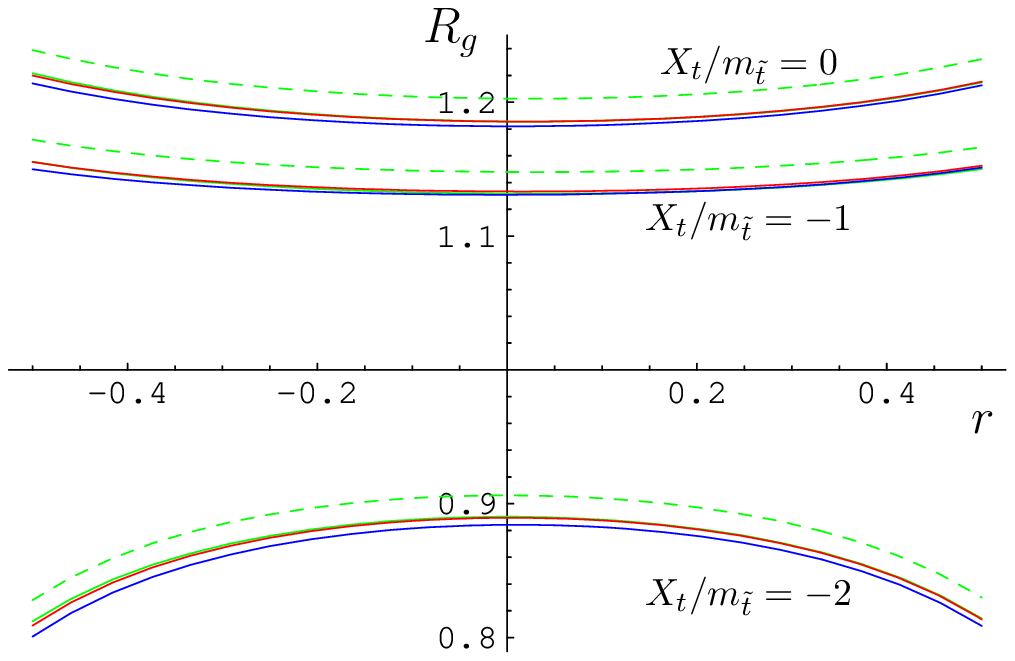}
\caption{Plot of $R_g$ as a function of 
$r$ for $m_{\tilde{t}}^2 = 500$ GeV,
$\tan \beta = 10$ (green/gray), $\tan \beta = 30$ (red/dark gray), $\tan \beta = 50$ (blue/black).
The solid lines are for other SUSY masses fixed to 400 GeV.
For comparison, the (green/gray) dashed 
lines are for other SUSY masses fixed to 800 GeV and $\tan \beta = 10$.
The three clusters of lines correspond to $X_t/m_{\tilde{t}} = 0,-1,-2$ as indicated in the plot.
}
\label{fig:hgg_and_r}
\end{figure}
If we further drop the subleading contribution proportional to $m_Z^2$ in $g_{h\tilde{t}_i\tilde{t}_i}$, 
then Eq.~(\ref{stophgg}) becomes
\begin{equation}
\label{stophgg1}
   \frac{m_t^2 (m_{\tilde{t}_1}^2 + m_{\tilde{t}_2}^2) +m_t^2 X_t^2}
        {\det M_{\tilde{t}}^2 }.
\end{equation}
Defining variables
\begin{equation}
\label{msdef}
m_{\tilde{t}}^2=\frac{m^2_{\tilde{t}_L}+ m^2_{\tilde{t}_R}}2, \quad 
r= \frac{m^2_{\tilde{t}_L}- m^2_{\tilde{t}_R}}{m^2_{\tilde{t}_L}+ m^2_{\tilde{t}_R}},
\end{equation}
we see that Eq.~(\ref{stophgg1}) depends mostly on $X_t$ and $m_{\tilde{t}}^2$, 
and weakly on $r$ which only
appears in the denominator. In Fig.~\ref{fig:hgg_and_r}
we demonstrate that $R_g$, the ratio of the full gluon fusion rate in the MSSM over the rate in the standard model, varies little for
$|r| \alt 0.4$. The value of $r = 0.4$ for $m_{\tilde{t}} = 500\ {\rm GeV}$ corresponds to $m_{\tilde{t}_L} \sim 590$ GeV 
and $m_{\tilde{t}_R} \sim 390$ GeV. Most SUSY breaking scenarios generate comparable $m_{\tilde{t}_L}$ 
and $m_{\tilde{t}_R}$, and since the renormalization group running of stop masses is dominated by the gluino mass,
the contribution of which is identical for both $m_{\tilde{t}_L}$ 
and $m_{\tilde{t}_R}$, the weak scale values of both masses remain close to each other. For example,
all the SPS benchmark scenarios for supersymmetry in Ref.~\cite{Allanach:2002nj}
have stop mass splittings that fall within $|r| \le 0.4$. Throughout this paper we use the publicly available
code {\tt FeynHiggs2.5} \cite{Heinemeyer:1998yj} to obtain numerical results presented in plots.
The set of relevant input parameters
we use throughout this study is the top quark pole mass at $m_t = 172.5$ GeV, bottom quark pole mass at $m_b=4.7$ GeV, and the pseudo-scalar
Higgs mass at $m_A=400$ GeV.
In Fig.~\ref{fig:hgg_and_r} we plot the production rate for $\tan\beta=10,\,30,\,50$ (although these three cases are plotted with different color/shades they are hard to distinguish because of the 
negligible dependence on $\tan \beta$)  and two different common 
masses of all other superpartners, 400 GeV (solid) and 800 GeV (dashed, only for $\tan \beta = 10$).
The three clusters of lines correspond to $X_t/m_{\tilde{t}} = 0,-1,-2$ as indicated in the plot.
It is clear that the dependence on $\tan \beta$ and masses of other superpartners is very small.

It is worth commenting that {\tt FeynHiggs} computes the approximate Higgs production cross-sections using 
extrapolation of the standard model production rate \cite{Hahn:2005cu}. Higher-order corrections such as the next-to-leading
order QCD corrections might be important in determining the Higgs production rate in the MSSM and should be
included in future analysis. 
However, the important observation relevant for our proposal is that the change in the gluon fusion 
production rate is largely a constant shift \cite{Harlander:2004tp} and does not introduce a significant dependence on other
SUSY parameters such as the gaugino mass $m_{1/2}$. In order to demonstrate our method we find it sufficient to use the 
approximation of {\tt FeynHiggs}.

\section{Stops and the Higgs mass}
\label{higgsmass}

In the Higgs sector of the MSSM there are two Higgs doublets, $H_u$ and $H_d$, 
coupling to the up-type and down-type
quarks respectively. After electroweak symmetry breaking three components are eaten
Goldstone bosons and give mass to the electroweak gauge bosons through the Higgs mechanism.
The remaining physical states are two CP-even neutral Higgs bosons, $h$ (the lighter one) and $H$
(the heavier one), one CP-odd neutral Higgs boson $A$, and the charged Higgses $H^\pm$. 
In the MSSM
supersymmetry imposes very tight constraints on the Higgs potential at tree-level, in particular,
the scalar quartic couplings are completely fixed by $SU(2)_w\times U(1)_Y$ gauge couplings.
As a result there are two free parameters in the MSSM Higgs sector, usually taken to be
$\tan\beta$ and $m_A$, and one can derive hierarchical relations for masses of 
different Higgs bosons~\cite{Djouadi:2005gj}. Among them the most important one is perhaps
the upper bound on the mass of the lightest Higgs boson $h$,
\begin{equation}
m_h \le m_Z |\cos 2\beta| \le m_Z = 91.2\ \  {\rm GeV} ,
\end{equation}
which is clearly below the LEP bound $m_h \ge 114$ GeV.

Therefore, one usually resorts to large radiative corrections from superpartners 
with significant coupling to the Higgs boson to raise $m_h$. 
This is why
$m_h$ is most sensitive to the parameters in the stop sector, and not to masses of other superparticles.~\footnote{The exception is again the sbottom sector for very large $\tan \beta$ and large $\mu$.}
For simplicity if we assume
$m_{\tilde{t}_R} \simeq m_{\tilde{t}_L} = m_{\tilde{t}}$, the
one-loop correction to $m_h$ is approximately given as
\begin{equation}
\Delta m_h^2 \simeq  \frac{3G_F}{\sqrt{2}\pi^2} m_t^4 \left\{ \log\frac{m_{\tilde{t}}^2}{m_t^2} + \frac{X_t^2}{m_{\tilde{t}}^2}
\left(1-\frac{X_t^2}{12 m_{\tilde{t}}^2} \right) \right\}  ,
\label{eq:mh}
\end{equation}
which grows logarithmically with the stop mass $m_{\tilde{t}}$. On the other hand,
the up-type Higgs mass squared parameter increases quadratically with $m_{\tilde{t}}$,
\begin{equation}
\Delta m_{H_u}^2 \simeq - \frac{3}{8\pi^2} m_{\tilde{t}}^2 \log \frac{\Lambda^2}{m_{\tilde{t}}^2}.
\end{equation}
It is the logarithmic versus quadratic dependence on the stop mass that dictates the fine-tuning
 in the MSSM. For $m_{\tilde{t}_R} \simeq m_{\tilde{t}_L}$ the stop masses need 
to be very large, 
${\cal O}$(1 TeV), to evade the LEP limit on the Higgs mass, which leads to large (${\cal O} (m_Z^2/m_{\tilde{t}}^2) \lesssim 1\%$) fine-tuning in 
electroweak symmetry breaking. 
On the other hand, the stop masses could be significantly below 1 TeV if there is large mixing in the stop sector,
in which case the fine tuning can be reduced to the level of 5\%.
The Higgs mass is maximized for $|X_t/m_{\tilde{t}}|\sim 2$ and with this mixing light stops, $m_{\tilde{t}_R} \simeq m_{\tilde{t}_L} \simeq 300$ GeV, are sufficient to push the Higgs mass above the LEP limit.

\begin{figure}
\includegraphics[scale=1.]{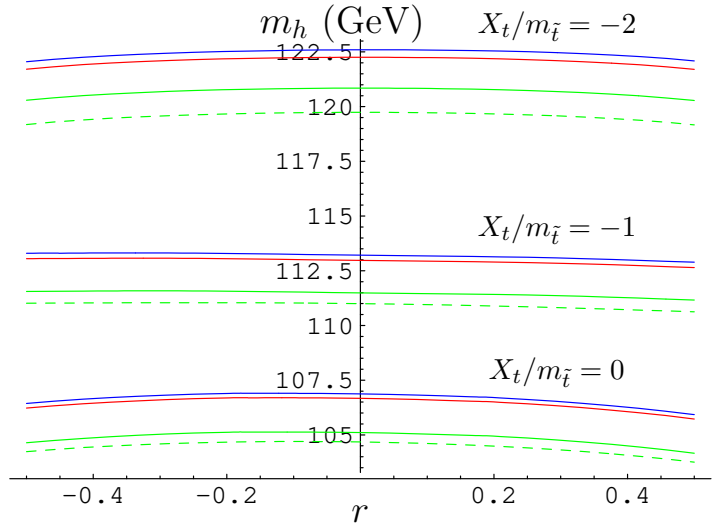}
\caption{Plot of  the Higgs boson mass as a function of 
$r$ for $m_{\tilde{t}}^2 = 500$ GeV,
$\tan \beta = 10$ (green/gray), $\tan \beta = 30$ (red/dark gray), $\tan \beta = 50$ (blue/black).
The solid lines are for other SUSY masses fixed to 400 GeV. For comparison, the (green/gray) dashed 
lines are for other SUSY masses fixed to 800 GeV and $\tan \beta = 10$.
The three clusters of lines correspond to $X_t/m_{\tilde{t}} = 0,-1,-2$ as indicated in the plot.}
\label{fig:mh_and_tanb}
\end{figure}

From the discussion above, we see that
for $\tan\beta\agt 10$ the tree level contribution to the Higgs mass is saturated and the residual dependence 
of the Higgs mass on   $\tan\beta$ is very weak ($\tan\beta$ does not enter the leading one loop correction).
Furthermore, in spite of Eq.~(\ref{eq:mh}) being derived for $m_{\tilde{t}_R} \simeq m_{\tilde{t}_L}$ and 
the Higgs mass in general being dependent on all three parameters in the stop sector:  $m_{\tilde{t}}$, $X_t$ and $r$, 
the dependence on $r$ is very weak for quite large deviations of  $m_{\tilde{t}_R}$ and $m_{\tilde{t}_L}$ from the average value. 
In Fig.~\ref{fig:mh_and_tanb} we plot the sensitivity of the Higgs mass to $r$
for three different values of $\tan\beta$ (distinguished by colors/shades) and two different common 
masses of all other superpartners, 400 GeV (solid) and 800 GeV (dashed, only for $\tan \beta = 10$).
The three clusters of lines correspond to $X_t/m_{\tilde{t}} = 0,-1,-2$ as indicated in the plot.
Again we see that $m_h$ is not much dependent of $r$, $\tan\beta$, and  masses of other superpartners
 in the region of the parameter space we are considering.

\section{Results}

\begin{figure}
\includegraphics[width=3in]{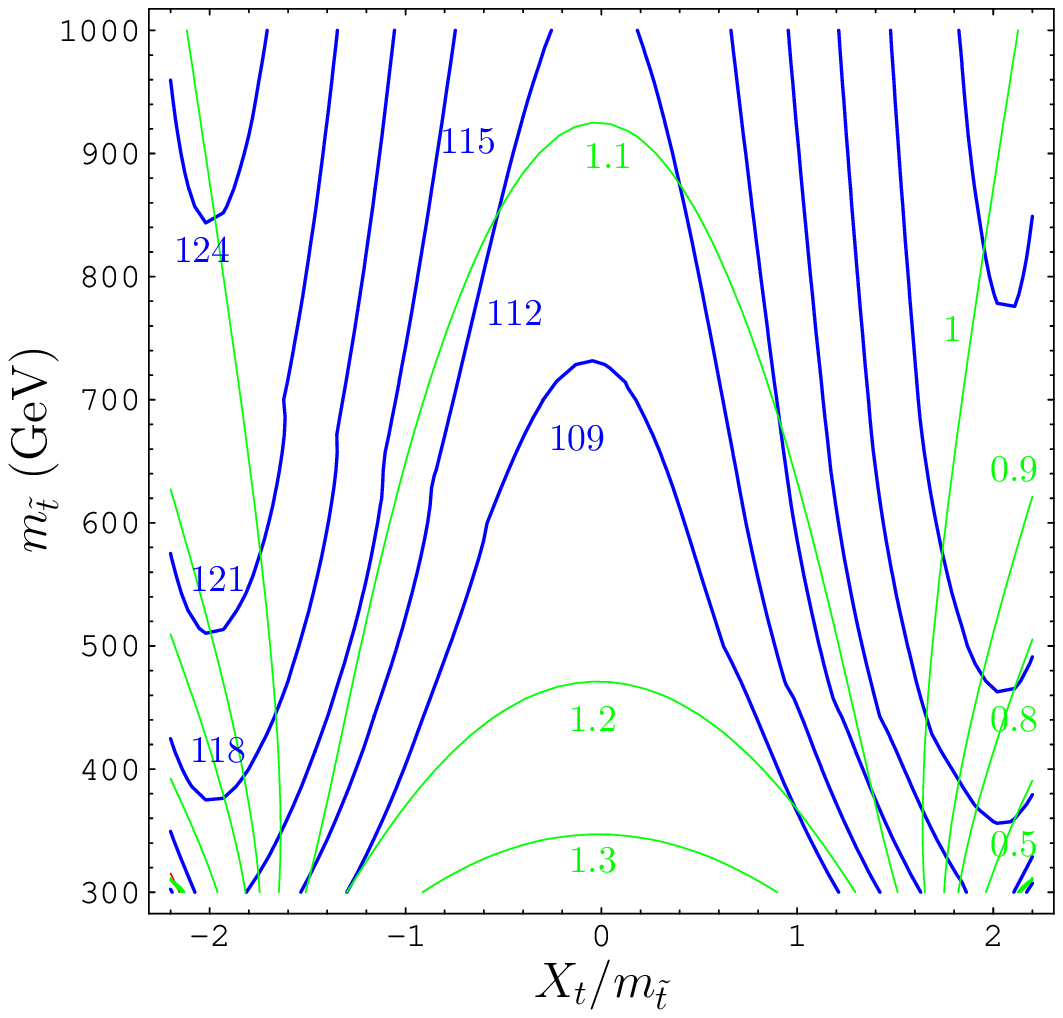}
\hspace{0.7cm}
\includegraphics[width=3in]{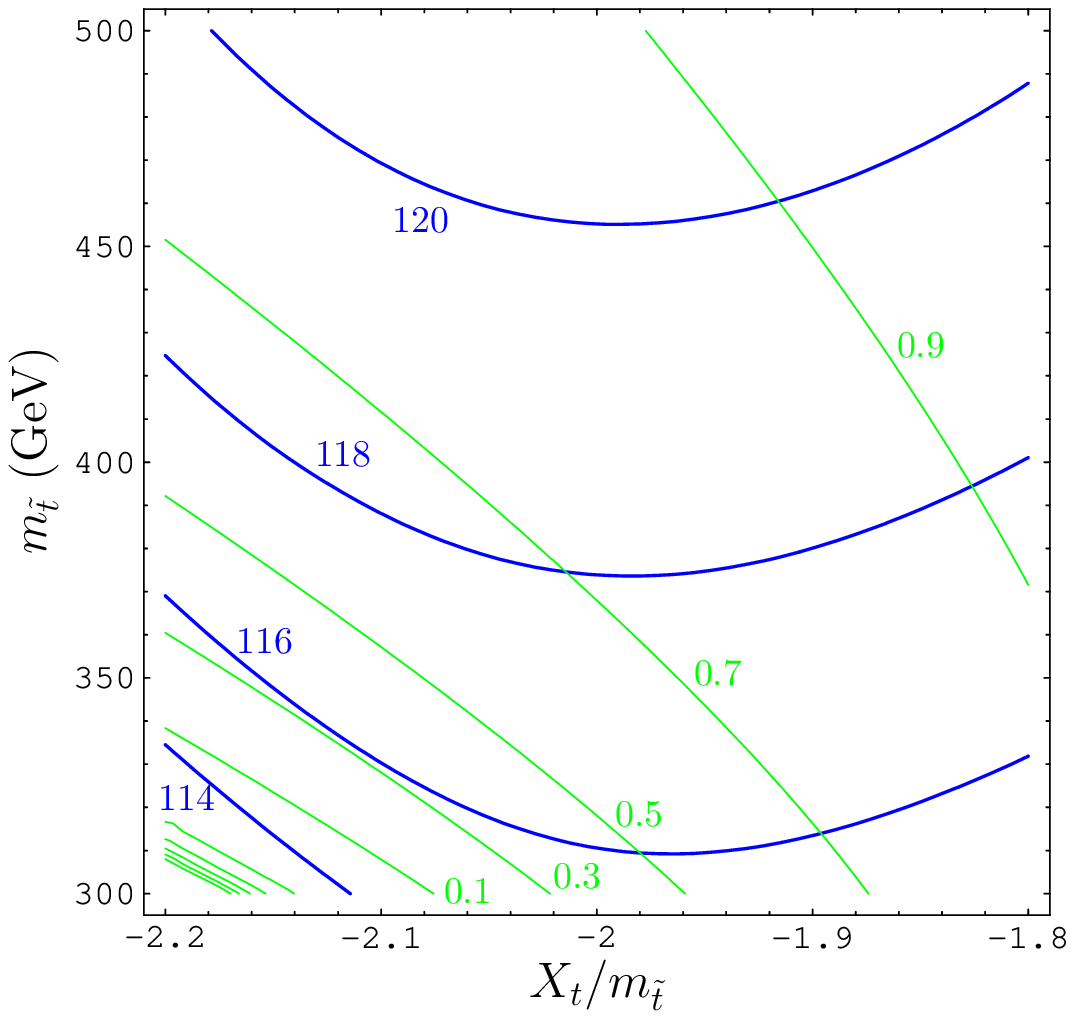}
\caption{Contours of constant Higgs mass $m_h$ (GeV) (blue/black) and the gluon fusion rate $R_g$ (green/gray) in 
$m_{\tilde t}$ -- $X_t/m_{\tilde t}$ plane. The plot on the right zooms in on the region of small $m_{\tilde t}$ and large mixing $X_t/m_{\tilde t}$. All other SUSY masses are fixed to 400 GeV, $\tan \beta = 10$ and $\mu = 200$ GeV.}
\label{fig:mh_and_hgg}
\end{figure}

In this section we present our results, concentrating on the observables $m_h$ and 
$R_g\equiv \Gamma_g^{\rm MSSM}/\Gamma_g^{\rm SM}$ which is the ratio of
the Higgs production rate in the gluon fusion channel in the MSSM and in the standard model. 
Contours of constant $m_h$ and $R_g$ are plotted in the $m_{\tilde{t}} - X_t/m_{\tilde{t}}$ plane, 
as shown in Fig.~\ref{fig:mh_and_hgg}.

Let us
focus first on the contours of constant $R_g$, observing that $R_g \agt 1$ when the mixing in the stop
sector is small $|X_t/m_{\tilde{t}}| \alt 1.6$ regardless of $m_{\tilde{t}}$. Moreover, for small mixing
$R_g$ increases as $m_{\tilde{t}}$
decreases, since lighter stops give more significant contributions to the production rate. On the other hand,
in the region where stops are light $m_{\tilde{t}} \sim {\cal O}$(500 GeV) and mixing is large 
$|X_t/m_{\tilde{t}}| \sim 2$, 
we see $R_g \alt 1$. The fact that the Higgs production in the gluon fusion channel in the MSSM could be 
smaller than 
in the standard model for large mixing in the stop sector
has previously been observed in Ref.~\cite{Djouadi:1998az}. It is interesting to note
that $R_g$ alone seems to give a good sense of the magnitude of $X_t/m_{\tilde{t}}$: $R_g \agt 1$ if 
the mixing is small and $R_g \alt 1$ if the mixing is large. 

For contours of constant Higgs mass, the story is similar to what has been said repeatedly in the literature. 
If there is no mixing in the stop sector, the stop mass $m_{\tilde{t}}$ needs to be close to 1 TeV in order 
to have a 
Higgs mass above the LEP bound of 114 GeV. The Higgs mass starts increasing when one turns on the
mixing and eventually reaches a maximum value for $|X_t/m_{\tilde{t}}| \sim 2$. In the region of large mixing  light stops, $m_{\tilde{t}} \simeq 300$ GeV, are  still allowed by  $m_h \ge 114$ GeV.

When we consider both kinds of contours together, there are several observations to be made. First consider
the region of small mixing. In this region contours of constant $m_h$ and $R_g$ run somewhat parallel to 
each other vertically, implying a very loose constraint on $m_{\tilde{t}}$, the overall stop mass scale,
unless the gluon production rate can be measured precisely in experiments. 
Furthermore, the region where $R_g \agt 1$
corresponds to the region where EWSB is more fine-tuned. Once we move into the region where 
$R_g \alt 1$, contours of constant $R_g$ run at large angles with contours of constant $m_h$, which means
it is possible to determine both $m_{\tilde{t}}$ and $X_t/m_{\tilde{t}}$ fairly well even if there is a large uncertainty 
in $R_g$. This is because in this region $R_g$ is quite sensitive to $m_{\tilde{t}}$ and (especially) 
$X_t/m_{\tilde{t}}$, and decreases rapidly with increasing mixing and decreasing stop masses. Therefore measurements
of $m_h$ and $R_g$ will allow for a fairly accurate determination of $m_{\tilde{t}}$ and $X_t/m_{\tilde{t}}$ in the region
of large mixing and light stops. All these measurements involve properties of the Higgs boson and can be done without
prior knowledge of other masses and mixing angles in the MSSM spectrum. 
As demonstrated in previous sections, these results are not sensitive to other parameters, and choosing different values of $\tan \beta$, $\mu$ or masses of other superpartners would only negligibly change results presented in  Fig.~\ref{fig:mh_and_hgg}.

In Fig.~\ref{fig:mh_and_hgg} we have also zoomed in on the region of large mixing with negative $X_t$ 
and small stop mass, since this is the region of particular interest: both $m_h$ and $R_g$ vary rapidly. 
This is also the least fine-tuned region of the MSSM. 
From the zoomed in plot one can see that, for example, if the experimental central values are $m_h=118$ GeV 
and $R_g=0.7$, then the corresponding central values for $m_{\tilde{t}}$ and $X_t/m_{\tilde{t}}$ 
will be 380 GeV and $-2$,
respectively. Of course we should not forget that there is another solution for roughly the same
$m_{\tilde{t}}$ but positive $X_t/m_{\tilde{t}}$. We also mention in passing that all the 
constant $R_g$ contours re-appear
in the dense region near the bottom-left corner \footnote{In fact, in this region $m_h < 114$ GeV which is ruled out by the LEP limit.} where 
$m_{\tilde{t}}\sim 300$ GeV and 
$X_t/m_{\tilde{t}}\sim -2.2$. 
They re-appear because in this region the lightest stop is extremely light, $\sim 120$ GeV,
for which the stop contribution in the gluon fusion rate, Eq.~(\ref{hggcorr}), completely overwhelms
the standard model contribution. Therefore the production rate at first decreases all the way to zero,
when stop contribution reaches a critical value and become equal and opposite to the top contribution,
and then starts growing when stop contribution becomes more negative than the critical value. In this
region $R_g$ is a very rapidly growing function for decreasing stop mass. The end result is a region
with very densely populated contours in the very bottom-left corner in Fig.~\ref{fig:mh_and_hgg}.

At this point we would like to comment on various theoretical and experimental uncertainties
one might encounter in implementing our strategy. For the lightest CP-even
Higgs boson $h$, if it is observed in the golden channel $h\to ZZ \to \ell^+\ell^-\ell^+\ell^-$ or the
silver mode $h\to \gamma\gamma$, then its mass can be measured with an accuracy of 
$\Delta m_h/m_h \sim 0.2\%$ at the LHC \cite{cmstdr}. For $m_h\sim 120$ GeV, this gives an uncertainty
of only 250 MeV! Unfortunately, the theoretical uncertainty in computing $m_h$ within MSSM is still
quite large in comparison. In the MSSM the full one-loop and dominant two-loop corrections to $m_h$ 
have been calculated, however, results from two different renormalization schemes differ by about
$2-3$ GeV \cite{Djouadi:2005gj}. The difference can be seen as a rough estimate of the 
magnitude of the unknown higher
order corrections. On the other hand, the situation with uncertainties
in the partial width $\Gamma(gg\to h)$, and hence $R_g$, is less optimistic. The reason is two-fold.
First the production rate of $gg \to h$ is not directly measurable in experiments since the Higgs
can only be seen through its decay products. Instead, what can be measured directly is
the cross-section times the branching ratio such as $\sigma(gg\to h)\times {\rm Br}(h\to 2\gamma)$.
By combining measurements on Higgs production and decay in different channels it is possible to extract
individual partial decay width, and at the LHC with a 200 fb$^{-1}$ luminosity 
the error is expected to be $\Delta \Gamma_g/\Gamma_g
\sim 30 \%$ when including systematic errors of approximately $20\%$ from higher order QCD 
corrections \cite{Zeppenfeld:2002ng,Belyaev:2002ua,Duhrssen:2004cv}. 

The uncertainty in the top quark mass has also an effect on calculation of both the Higgs mass and the gluon 
fusion production of the Higgs boson. At the LHC the mass of the top quark is expected to be measured with 
uncertainties of 1 GeV, dominated by systematic errors~\cite{Borjanovic:2004ce,Etienvre:2006ph}. 
An uncertainty at this level results in $\sim 0.5$ GeV difference in the calculated Higgs mass which is much 
smaller compared to the theoretical uncertainty. For the gluon fusion production 
the top quark mass uncertainty is negligible compared to the systematic error discussed above.
Furthermore, a recent study
suggests that a significantly better precision of the top quark mass measurement can be achieved
using a sequence of effective field theories for the reconstruction of the top quark invariant mass distributions 
at collider experiments~\cite{Fleming:2007qr}.

In the end, the uncertainty in $m_h$ is expected to be at the level of 2$\%$, dominated completely by
theoretical uncertainty, whereas the uncertainty  
in $R_g$ is much larger, at the level of 30$\%$. However, we should stress that, even with
a 30$\%$ uncertainty in $R_g$, in the region of large mixing and small stop mass it
could still be useful to apply our strategy due to the fact that $R_g$ is very sensitive
to $m_{\tilde{t}}$ and $X_t$ in this region. For example, even if the production rate is poorly measured
to be in the region $0.7 \agt R_g \agt 0.3$, it is still possible to constrain the $m_{\tilde{t}} - X_t/m_{\tilde{t}}$
plane down to a small area by knowing $m_h$ with 2 GeV uncertainty, as can be seen
from Fig.~\ref{fig:mh_and_hgg}.

\section{Extreme Corners of the MSSM}

In this section we use our results to explore an interesting possibility 
that measurements of the Higgs mass and the 
production rate do not have overlapping contours in the $m_{\tilde{t}} - X_t/m_{\tilde{t}}$ plane.
Both $m_h$ and $R_g$ is a measure on the overall stop mass scale $m_{\tilde{t}}$ and the mixing
$X_t$. If $m_{\tilde{t}}$ and $X_t$ inferred separately from $m_h$
and $R_g$ are very far off, then it is a signal that the region of parameter space we
are considering,
\begin{equation}
10 \alt \tan\beta \alt m_t/m_b,\quad 
m_b \, |\mu \tan \beta| \; \lesssim \; m^2_{\tilde{b}_L}, \ m^2_{\tilde{b}_R},
  \quad {\rm and} \quad |r| \alt 0.4,
\label{eq:sanity}
\end{equation}
 is disfavored. In this case, we can further ask if it is possible to reconcile the
differences in these two measurements by considering other parameter regions.

From Fig.~\ref{fig:mh_and_hgg} we see that the only situation in which contours from measurements 
of the Higgs mass and the production rate would not overlap (taking into account uncertainties discussed in the previous section) is when the Higgs is relatively heavy,
$m_h \agt 130$ GeV, and production rate very small, $R_g \alt 0.6$. The reason is
that a Higgs mass around 130 GeV requires a high stop mass scale, $m_{\tilde{t}} \agt 1$ 
TeV, whereas a small production rate prefers a low stop mass scale, $m_{\tilde{t}} \alt 500$ GeV.

In order to find a resolution in these two measurements, it is necessary to find ways to 
lower $m_{\tilde{t}}$ while keeping $m_h$ fixed at around 130 GeV, or increase $m_{\tilde{t}}$
while maintaining a small $R_g$ at roughly 0.6. Immediately we conclude that going to a smaller
$\tan\beta$ would not help because, in this case, the tree-level contribution to the Higgs
mass is reduced and $m_{\tilde{t}}$ has to be even higher in order to produce a larger
radiative corrections to keep $m_h$ large. This worsens the discrepancy. 

An alternative is
to have a large $\tan\beta \sim  m_t/m_b$ for which the sbottom contributions are important.
Let us first discuss the effect of the sbottom sector on both the Higgs mass and 
the production rate. In the decoupling limit, a large $\tan\beta$ is only a necessary
but not a sufficient condition for the sbottom effects to be important for the Higgs mass;
a sizable $\mu$ term is also required. In this case the mixing in the sbottom sector can
be very large which
has a tendency to 
decrease the Higgs mass \cite{Brignole:2002bz}. Obviously larger
$\mu$ term causes even larger $X_b$ and therefore smaller Higgs mass. Moreover the Higgs
mass is not an even function in $X_b \to -X_b$ and hence not in $\mu \to -\mu$ either.\footnote{
Notice that our definition of $\mu$ differs from that in \cite{Brignole:2002bz} by a sign.}
Since the sbottom mass matrix and sbottom
couplings to the lightest CP-even Higgs are very similar to those in the stop sector given in Sect.~\ref{glue}
with the corresponding electroweak charges, masses, and mixing term replaced by those for the
sbottom, we expect that, in the same fashion as the stop, if the sbottom is light
and mixing is large, it could decrease the production of the Higgs in the gluon fusion channel.
The production rate does not depend explicitly on the sign of $X_b$ but only implicitly through
the Higgs mass $m_h$.

Now in order to produce a large effect in the production rate the sbottom has to be light,
since its contribution decouples as $1/m_{\tilde{b}}^2$.
On the other hand, the stop must be heavy to keep the Higgs mass large.
At this point it is important to keep in mind that the soft-breaking masses for the
left-handed sfermions are required to be the same by the $SU(2)_w$ gauge symmetry:
$m_{\tilde{t}_L}^2 = m_{\tilde{b}_L}^2= m_{\tilde{q}_3}^2$. Therefore there is a limited number of ways to
keep at least one of the sbottoms light and at least one of the stops heavy.
As an example, $m_h \sim 130$ GeV and $R_g \sim 0.6$ can be produced with
the following choices of parameters (assuming large mixing
in the stop sector that maximizes $m_h$): 

(a)  $\tan\beta \sim 50$,  $m_{\tilde{q}_3} \sim m_{\tilde{t}_R} \sim 2000$ GeV, 
     $m_{\tilde{b}_R} \sim 100$ GeV, and $\mu \sim -800$ GeV,\footnote{In this case
    the lightest sbottom $\tilde{b}_1$ is slightly lighter than 100 GeV. However, if
    $\tilde{b}_1$ is mostly right-handed, which is the case here, the limit on its
    mass is very weak, much lower than 100 GeV \cite{Carena:2000ka}.} 

(b) $\tan\beta \sim 50$,  $m_{\tilde{q}_3} \sim m_{\tilde{b}_R} \sim 300$ GeV, 
     $m_{\tilde{t}_R} \sim 5000$ GeV, and $\mu \sim -250$ GeV,

\noindent and small variations of these. As one can see, reconciling these two measurement in 
$m_h$ and $R_g$, by going outside of the choices of parameters we considered in Eq.~(\ref{eq:sanity}),
would require huge hierarchies in and between the stop and sbottom sectors.
Such hierarchies are difficult to  generate from a sensible UV model and  we consider them rather extreme. 
The more plausible explanation of conflicting values of $m_h$ and $R_g$ would be 
contributions from physics beyond the MSSM.

\section{Discussion and Conclusions}

In this paper we proposed using the Higgs boson as a probe of the stop sector.
Our method relies on
measurements of the Higgs mass as well as the production rate in the
gluon fusion channel, the dominant production channel at the LHC. For
 $m_t/m_b \agt \tan\beta \agt 10$ and small $\mu$ term, our proposal
is insensitive to other mass parameters in the MSSM and thus
complementary to the conventional method of studying the production and decay processes of 
stops, which requires knowledge of masses and mixing angles in the chargino
and neutralino sector.

In the stop mass-squared matrix, there are three free parameters $m_{\tilde{t}_L}^2,
m_{\tilde{t}_R}^2$ and $X_t$ which (roughly) correspond to the two diagonal entries and
the one off-diagonal entry. With only two measurements, the Higgs mass and the production rate in 
the gluon fusion channel, one might expect that a priori it is only possible to constrain the three parameters
on a one-dimensional surface. Nevertheless, we demonstrated that both measurements are sensitive
to only two out of the three parameters in the mass matrix; there is a (almost) flat direction 
in the space of parameters. In the end, two measurements provide an access to, in terms of variables
defined in Eq.~(\ref{msdef}), the overall stop mass scale
$m_{\tilde{t}}^2$ and the mixing term $X_t$, as long as $|r|\le 0.4$.
It is worth pointing out that all the Snowmass benchmark scenarios for the MSSM have mass
splittings satisfying $|r|\le 0.4$.  We also note that
very often $r$ is calculable from a given UV model in which case it is not a free parameter and 
 our procedure can be used to determine the stop sector of the model completely.

The proposed strategy is the least effective when the mixing in the stop sector is not large, for
in this region contours of two different measurements run in parallel to each other. This
happens when the Higgs is light and the production rate is close to the standard model value.
On the other hand, our method is the most effective when stops are light and the mixing is large, 
in which case 
the allowed area in the $m_{\tilde{t}} - X_t/m_{\tilde{t}}$ plane is quite small. Because 
the production rate
is very sensitive to $m_{\tilde{t}}$ and $X_t$ in this particular region, even with an uncertainty as 
large as $30\%$ in the production rate, our proposal could be useful as discussed in the
previous section.

As already emphasized, our proposal should be considered as complementary to methods of extracting stop 
mass parameters in direct production and decay processes. The point is to measure the same set of 
parameters in as many different ways as possible and see if there is a consistent set of numbers emerging.
The computation presented in this study is at best exploratory in nature, since it does not include many of 
the recent higher order calculations of the Higgs production and decay rates. We only wish to demonstrate the 
feasibility of the proposal and identify regions of parameter space where the method is the most effective, in
order to motivate and facilitate future studies. 

We also considered the case when there is a discrepancy between the measurements of the Higgs mass and
the production rate. This could happen if the lightest CP-even Higgs is heavy, $m_h \gtrsim 130$ GeV, and the production
rate is significantly smaller than in the standard model. Even though it is possible to generate such
a pattern in the MSSM, the required spectrum has large mass hierarchies in and between the stop and sbottom 
sectors which resides in extreme corners of the parameter space.

As a final comment, the effectiveness of our strategy is clearly limited by the possibly 
large uncertainty incurred in the measurement of the production rate in the gluon fusion channel.
We hope our proposal could serve as a strong motivation to make an effort to
reduce the uncertainty in the gluon fusion production rate, either through a better
experimental measurement or a more precise theoretical calculation. Before that
goal is achieved, a better observable to consider is probably the event rate of 
$gg\to h \to \gamma\gamma$, which is 
directly measurable and has less uncertainty. However, the decay rate to two photons
in the MSSM depends not just on stop masses but also on  chargino masses. Thus if charginos are
observed at the LHC and their masses are measured, then it could be useful to combine the 
measurement of $gg\to h \to \gamma\gamma$ with the Higgs mass, 
in the same fashion as described in this paper, 
to constrain the stop sector of the MSSM.

\begin{acknowledgments}
We thank Thomas Hahn and Sven Heinemeyer for help and correspondences
on {\tt FeynHiggs}. RD thanks Hyung Do Kim for useful discussions.
This work is supported in part by the 
National Science Foundation under grant PHY-0653656 and the 
Department of Energy under grant DE-FG02-90ER40542.
\end{acknowledgments}




\end{document}